\documentclass[12pt]{article}
\usepackage{graphicx}
\usepackage{amssymb}
\usepackage{epstopdf}
\usepackage{color}
\def\beq{\begin{equation}}
\def\eeq{\end{equation}}
\def\bea{\begin{eqnarray}}
\def\eea{\end{eqnarray}}
\def\xitime{\xi_{\rm time}}
\def\xirot{\xi_{\rm rot}}
\def\dtdtau{\frac{dt}{d\tau}}
\def\drdtau{\frac{dr}{d\tau}}
\def\dphidtau{\frac{d\phi}{d\tau}}

\def\bnotes{\noindent {\it Comments for Further Consideration in this Section}\begin{itemize}}
\def\enotes{\end{itemize}}
\def\bvec#1{\mathbf{#1}}

\def\rbob{R_{\rm bob}}
\def\ralice{R_{\rm alice}}
\def\vbob{v_{\rm bob}}

\begin{document}


\title{\bf\large When effective theories predict: the inevitability \\ of Mercury's anomalous perihelion precession}
\author{James D. Wells\\
{\it CERN Theoretical Physics (PH-TH), CH-1211 Geneva 23, Switzerland} \\
{\it Department of Physics, University of Michigan, Ann Arbor, MI 48109}}
\date{\today}

\maketitle

\begin{abstract}

If the concepts underlying Effective Theory were appreciated from the earliest days of Newtonian gravity, Le Verrier's announcement in 1845 of the anomalous perihelion precession of Mercury would have been no surprise. Furthermore, the size of the effect could have been anticipated through ``naturalness" arguments well before the definitive computation in General Relativity. Thus, we have an illustration of how Effective Theory concepts can guide us in extending our knowledge to ``new physics", and not just in how to reduce larger theories to restricted (e.g., lower energy) domains.

\end{abstract}

\vspace{0.1in}

\begin{center}
{\it Lectures presented to Philosophy of Physics \\ students at the University of Michigan}
\end{center}

\vfill\eject

\tableofcontents

\vfill\eject

\section{Introduction}

The purpose of these lectures is to introduce the concepts of Effective Theories to students of Philosophy, Mathematics and Physics who have a shared interest in the philosophy and history of physics. The concept I wish to discuss, Effective Theory, is a thoroughly modern notion; nevertheless, I wish to illustrate it with a very old and intuitively accessible problem in physics: Mercury's anomalous perihelion precession.  

Le Verrier announced in 1845 a small discrepancy in the precession rate of Mercury's perihelion compared to Newton's theory, even after taking into account all the disturbing influences throughout the solar system such as the effect of other planets' orbits\footnote{In 1859 Le Verrier gave a number for this advance: 35 arcseconds per century~\cite{leverrier:1859}. It was later reevaluated by S. Newcomb~\cite{Newcomb1882}, who determined the correct value of 43 arcseconds per century.}.  This came as a surprise, and more or less nobody believed at the time that it was the fault of Newton, but rather the fault of observers who had not seen the other celestial bodies that must surely be perturbing Mercury's orbit.  Historically, that is the beginning of the problem. Le Verrier believed that an as-yet unobserved mass distribution inside the orbit of Mercury was the source of the discrepancy. He and others advocated the existence, for example, of a new small planet  (``Vulcan" as it was sometimes called) that would be observed when astronomers developed the instruments necessary to find it~\cite{Roseveare:1982}. Such was not the case.
By the 1890's it became clear to most that new large-scale object(s) was not the explanation~\cite{Oppenheim:1920}, despite some ill-fated protestations otherwise~\cite{Poor:1921}. The resolution of the problem came with Einstein's General Relativity, which predicted precisely the 43'' of arc per century observed, and the case was closed.

However, I want to argue that anticipation of the ``problem" {could have} occurred much before Le Verrier.  What prevented scientists from anticipating Mercury's perihelion precession was not lack of mathematical skill, or lack of experimental abilities.  It was solely due to not having the right {mindset}.  Unlike perhaps in decades and centuries gone by, no competent scientist should retain an unfailing commitment to any theory. {All theories are incomplete, even given that some theories are better than others}.  The code phrase of this mindset is {Effective Theories}. The concept is a powerful one that has born much fruit in theories of particle physics,  condensed matter systems, and even cosmology.  

These notes are meant to be a somewhat pedagogical and technical exposition of the Mercury problem and the application of Effective Theory ideas to the problem.  In some parts of this lecture I will follow an ``alternative history" path with the scientists Alice and Bob who vaguely understand the importance of Effective Theories and who will devise a theory that can accomodate the perihelion precession rate well before Einstein's General Relativity comes along, and may even be able to predict roughly the numerical rate of the precision and make predictions for other planets through ``naturalness" arguments.  The latter could have been possible after  diligent reflections on the philosophical challenges of Newton's theory.  I will compute the General Relativity rate at the end, in order to show how elegantly it comes out of that more complete theory, and to show that it matches the Effective Theory ``predictions" by Bob and Alice. And finally I will conclude with some more remarks on the meaning of the results.

\section{Orbits in Newton's Theory}

To remind some students who have not seen celestial mechanics for some time,
we begin with the computation of particle orbits in Newton's gravity.  The reader familiar with these basics should feel free to skim the section only for definitions and conventions that I will use later.

We know that the orbits predicted by Newton's law of gravity are respected quite well by the planets, and so any change in the equations of motion for the orbits will need to be small perturbations.
In Newton's gravity, a test particle with mass $m$ orbits around a particle of mass $M\gg m$ according to the equations of motion derived from the lagrangian
\beq
L=\frac{1}{2}m\dot \bvec r^2+\frac{\alpha}{r}
\eeq
where $\alpha=GMm$ with $G$ being Newton's constant.  $M$ represents the sun's mass in this lecture, and $m$ the planet's mass, most often Mercury. It is appropriate to assume that  $M\gg m$ such that any correction is negligible due to the difference of $m$ from the reduced mass $\mu=Mm/(M+m)$, which is technically the precise mass one should use in the kinetic energy term.  Lagrange's equations of motion are
\beq
m\ddot\bvec r=-\frac{\alpha}{r^2}\hat\bvec r
\eeq
where $\hat\bvec r$ is the unit vector in the $\bvec r$ direction.

\subsection{Orbital Solution}

The lagrangian is rotationally invariant, and so the motion of the particle  is  most conveniently evaluated by casting the vector equation of motion into the two polar component equations
\bea
m(\ddot r-r\dot \phi^2)& = & -\frac{\alpha}{r^2}~~{\rm (radial~equation)}\\
m(2\dot r\dot \phi +r\ddot \phi) & = & 0~~{\rm (angular~equation).}
\eea
The second equation is equivalent to 
\beq
\frac{d}{dt}(mr^2\dot \phi)=0
\label{eq:angular}
\eeq
which implies that $mr^2\dot\phi$ is a constant in time.
At the apogee (furthest) or perigee (closest) point of the orbit the radius vector $\hat\bvec r$ is exactly perpendicular to the angular vector $\hat\bvec\phi$ and the magnitude of the angular momentum vector $\vec \ell=\bvec r\times \bvec p$, where $\bvec p=mr\dot \phi\hat\bvec\phi$, is exactly $mr^2\dot\phi$. Since angular momentum is conserved and $mr^2\dot\phi$ is conserved, if they are equal at one point they are equal at all points in the orbit. Thus, the constant value inside the time derivative eq.~(\ref{eq:angular}) is none other than angular momentum: $\ell=mr^2\dot\phi$. This also proves that the motion is in a plane. Since angular momentum is a conserved vector quantity, the direction must also be preserved which is only possible if $\bvec p$ perpetually lies in the same plane as $\bvec r$.
This justifies our evaluation of a three-dimensional problem in terms of just two variables $(r,\phi)$ in the plane of motion.

Let us now solve the radial differential equation to obtain an exact solution of the orbit for particle $m$.   By rewriting $r\equiv 1/u$, recasting all time derivatives as $d/dt\to \dot \phi d/d\phi$ when possible, and recognizing that $\dot\phi = l/r^2m$ from conservation of angular momentum, one finds that the governing differential equation of motion is
\beq
\frac{d^2u}{d\phi^2}+u=\frac{\alpha m}{\ell^2}.
\eeq
Interestingly, this equation takes the form of the harmonic oscillator equation. The solution is
\beq
u(\phi)=u_0\cos\phi +\frac{\alpha m}{\ell^2},
\eeq
where $u_0$ is a constant that is not determined by the theory but the particular circumstances (i.e., initial conditions) of the system.  In terms of the more direct variable $r$, the solution is
\beq
r(\phi)=\frac{\rho}{1+e\cos\phi}, ~{\rm where}~\rho=\frac{\ell^2}{\alpha m},~{\rm and}~e=u_0\rho,
\eeq
and it is assumed that $\phi=0$ is at perigee. $\rho$ is sometimes called the lactus rectum of the orbit.

The constant $e$ is called the eccentricity with which one can classify an orbit  as circular ($e=0$), elliptical ($0<e<1$), parabolic ($e=1$), or hyperbolic ($e>1$). Focusing on the $0\leq e<1$ case of elliptical or circular orbits, we find that 
\beq
r_{\rm min}=\frac{\rho}{1+e},~~{\rm and}~~r_{\rm max}=\frac{\rho}{1-e}.
\eeq
The relation between the semimajor axis $a$ of the elliptical orbit and the other variables is given by
\beq
a=\frac{r_{\rm min}+r_{\rm max}}{2} ~~~{\rm which~implies}~~~\rho=a(1-e^2).
\eeq

\subsection{The Hamiltonian and $V_{eff}$ Description}

An alternative way to approach the problem is to compute the Hamiltonian and consider the orbit from the perspective of a one-dimensional effective potential for radial motion.  I provide the very basics of this to remind the students of the formalism which is used by some papers relevant to the perihelion precession. We first expand the lagrangian in terms of radial  and angular coordinates starting from the identity 
\bea
\dot {\bf r}^2 &= &\dot r^2+r^2\sin^2\theta\dot\phi^2+r^2\dot\theta^2 \nonumber \\
 & \Longrightarrow & \dot r^2+r^2\dot\phi^2~{\rm (valid~in~the~}\sin\theta=1{\rm ~fixed~orbital~plane)}
\eea
The Hamiltonian is constructed as
\beq
H=\sum_i \dot q_i p_i-L
\eeq
using the momentum factors
\beq
p_r=\frac{\partial L}{\partial \dot r}=m\dot r,~~{\rm and}~~p_\phi=\frac{\partial L}{\partial \dot\phi}=mr^2\dot \phi,
\eeq
which implies
\beq
H=\frac{p_r^2}{2m}+\frac{p_\phi^2}{2mr^2}-\frac{\alpha}{r}.
\label{eq:Hamiltoniana}
\eeq
The Hamiltonian is independent of $\phi$, which implies from Hamilton's equations of motion, 
\beq
\dot p=-\frac{\partial H}{\partial q},~~~{\rm and}~~~\dot q=\frac{\partial H}{\partial p},
\eeq
that $\dot p_\phi$ is a conserved quantity:
\beq
\dot p_\phi=\frac{\partial H}{\partial \phi}=0~~\Longrightarrow~~p_\phi={\rm const}.
\eeq
This of course is just a restatement of the conservation of angular momentum
\beq
\ell=p_\phi=mr^2\dot\phi .
\label{eq:angmom}
\eeq

We can substitute~eq.~\ref{eq:angmom} back into eq.~\ref{eq:Hamiltoniana}, which gives a one-dimensional Hamiltonian as promised:
\beq
H=\frac{p_r^2}{2m}+\frac{\ell^2}{2mr^2}-\frac{\alpha}{r}.
\eeq
The Hamiltonian is a constant of the motion -- the energy of the system -- and it is useful sometimes to consider the dynamics of particle motion from this consideration where $E\equiv H=T+V$ and $T$ and $V$ are the kinetic and potential energies respectively. Here the potential for the one dimensional motion is often called the effective potential and is given by
\beq
V_{eff}(r)=\frac{\ell^2}{2mr^2}-\frac{\alpha}{r}.
\eeq
It is this potential that governs the radial potential with the first term pushing the particle away from the origin and the second term attracting the particle to the origin. The balance giving orbital motion between two turning points of zero radial kinetic energy, the apogee and perigee.

\section{Perihelion precessions from perturbations}

From the previous section we know that the orbit from Newton's simple  $1/r^2$ force law is
\beq
u(\phi)=\frac{1}{r(\phi)}=\frac{1}{\rho}(1+e\cos\phi).
\eeq
This obviously does not allow any advancement of the perihelion. The minimum is where $du/d\phi =0$, which implies $\sin\phi=0$ and therefore $\phi=0,2\pi,4\pi,\ldots$ mark the successive perihelions.
The discovery of the anomalous perihelion precession of Mercury, if it can be established, would signal the end of the Newtonian era and initiate the search for a better theory.

\subsection{$1/r^2$ Correction to the Central Potential}

Let us look at how the orbits change if we add a $1/r^2$ correction to the potential of the gravitational interaction lagrangian.   Let us call this Bob's theory with lagrangian
\beq
L =  \frac{1}{2}m\dot\bvec r^2+\frac{\alpha}{r}\left( 1+\frac{\rbob}{r}\right) \\
\eeq
where $\alpha=GMm$, with $G$ being Newton's constant, $M$ is the mass of the sun, and $m$ is the mass of the planet under consideration.  This new law requires the introduction of a new fundamental length scale $\rbob$, which is a priori unknown. However, we do know, as will be shown below, that it leads to a perihelion precession of the orbits governed by this law.

Lagrange's equation of motion for this theory is
\bea
{\rm radial} & : & m(\ddot r-r\dot\phi^2)=-\frac{\alpha}{r^2}\left( 1+\frac{2\rbob}{r}\right) \\
{\rm angular} & : & m(2\dot r\dot\phi +r\ddot \phi)=0
\eea
The angular equation yields conservation of angular moment $\ell =mr^2\dot \phi={\rm const}$ just as before. Using this, we can write the radial differential equation as
\beq
\frac{d^2u}{d\phi^2}+u=\frac{\alpha m}{\ell^2}(1+2\rbob u)
\eeq
This can be rewritten as
\beq
\frac{d^2u}{d\phi^2}+\left( 1-\frac{2\rbob}{\rho}\right) u=\frac{1}{\rho},~~{\rm where}~~
\rho=\frac{\ell^2}{\alpha m}.
\eeq
The general solution to this equation, assuming perihelion is placed at $\phi=0$, is
\beq
u(\phi)=u_0\cos\left( \phi\sqrt{1-\frac{2\rbob}{\rho}}\right)+\frac{1}{\rho-2\rbob},
\eeq
or, written differently,
\beq
u(\phi)=\left(\frac{1}{\rho -2\rbob}\right)\left[ e\cos\left( \phi\sqrt{1-\frac{2\rbob}{\rho}}\right)+1\right]
\label{eq:u bob}
\eeq
where $e=u_0(\rho-2R)$.

The $u(\phi)$ solution describes the motion of a precessing ellipse. The first perihelion by definition is at $\phi=0$ and the second perihelion occurs when
\beq
\phi \sqrt{1-\frac{2\rbob}{\rho}}=2\pi~~\Longrightarrow~~
\phi=\frac{2\pi}{ \sqrt{1-\frac{2\rbob}{\rho}}}=2\pi+{2\pi}\frac{\rbob}{\rho}
\label{eq:deltarbob}
\eeq
The small perihelion advance is the deviation of $\phi$ from $2\pi$ and is $\delta =2\pi\rbob/\rho$.

Given our previous computations, we are now able to evaluate the relationship between the extra length scale $\rbob$ and the perihelion advance of an orbit.  In one case, if we have made a measurement of the perihelion advance, we can derive what value $\rbob$ must be to reproduce that value
\bea
\rbob  & = & (1.16\, {\rm km}) \left( \frac{\delta/T_{\rm orbit}}{{\rm arcsec}\cdot {\rm century}^{-1}}\right)\left( \frac{\rho}{{\rm 1\, au}}\right) \left( \frac{T_{\rm orbit}}{{\rm 1\, year}}\right)
\eea
where $\rho$ is related to the common parameters of the semimajor axis $a=(r_{\rm min}+r_{\rm max})/2$ and eccentricity $e$ by the relation $\rho=a(1-e^2)$.

On the other hand, if we have a theory for what $\rbob$ should be, we can make a prediction for the perihelion advance in units of arc seconds per century:
\beq
\frac{\delta}{T_{\rm orbit}} = \frac{2\pi\rbob}{\rho T_{\rm orbit}}= (0.866\, {\rm arcsec}\cdot {\rm century}^{-1})
\left( \frac{1\, {\rm au}}{\rho}\right)\left( \frac{\rm years}{T_{\rm orbit}}\right)
\left( \frac{\rbob}{1\, {\rm km}}\right)
\label{deltatorbitbob}
\eeq

\begin{table}[t] 
\centering 
\begin{tabular}{ccccccc}
\hline\hline
Planet & $T_{\rm orbit}$ (yrs) &  $e$ & $a$ (au)  & $\rho$ (au) & $r_{\rm min}$ (au) & $r_{\rm max}$ (au) \\
\hline 
Mercury & 0.241  & 0.206 & 0.387 &   0.371 & 0.307 & 0.467 \\
Venus & 0.615  & 0.007 & 0.723 & 0.723 & 0.718 & 0.728 \\
Earth & 1.000  & 0.017 & 1.000 & 1.000 & 0.983 & 1.017 \\
Mars & 1.881  & 0.093 & 1.524 & 1.511 & 1.382 & 1.666 \\
Jupiter & 11.86 & 0.048 & 5.203 & 5.191 & 4.953 & 5.453 \\
Saturn & 29.46  & 0.056 & 9.539 & 9.509 & 9.005 & 10.07 \\
Uranus & 84.02  & 0.047 & 19.19 & 19.15 & 18.29 & 20.09 \\
Neptune & 164.8  & 0.009 & 30.06 & 30.06 & 29.79 & 30.33 \\
Pluto & 247.7  & 0.249 & 39.46 & 37.01 & 29.63 & 49.29 \\
\hline\hline
\end{tabular} \caption{Data for planetary orbits. $T_{\rm orbit}$ is the time for one full revolution in earth years, $e$ is the eccentricity of the orbit, $a$ is the semimajor axis in astronomical units ($1\, {\rm  au} =1.496\times 10^{11}\, m$),  $\rho=\ell^2/GM m^2=a(1-e^2)$ is the orbital latus rectum in astronomical units (and is independent of $m$ ultimately), $r_{\rm min}=a(1-e)$ is the distance of perigee in astronomical units, and $r_{\rm max}=a(1+e)$ is the distance of apogee in astronomical units.} \label{orbit data} 
\end{table}

\subsection{$1/r^3$ Correction to the Central Potential}

Alice's theory has a $1/r^3$ correction to the potential
\bea
L_{\rm alice} & = &  \frac{1}{2}m\dot\bvec r^2+\frac{\alpha}{r}\left( 1+\frac{\ralice^2}{r^2}\right),
\eea
which gives a $1/r^4$ correction to the gravitational force law. Lagrange's equations for her theory are
\bea
{\rm radial} & : & m(\ddot r-r\dot\phi^2)=-\frac{\alpha}{r^2}\left( 1+\frac{3\ralice^2}{r^2}\right) \\
{\rm angular} & : & m(2\dot r\dot\phi +r\ddot \phi)=0
\eea
Here again the angular momentum $\ell =mr^2\dot \phi$ is conserved from the angular equation, and the radial equation becomes
\beq
\frac{d^2u}{d\phi^2}+u=\frac{\alpha m}{\ell^2}(1+3\ralice^2 u^2).
\label{eq:u alice}
\eeq

We'll solve this equation employing techniques of perturbation theory. We treat the last term of eq.~\ref{eq:u alice} as a small perturbation and solve first the equation
\beq
\frac{d^2u}{d\phi^2}+u=\frac{\alpha m}{\ell^2}
\eeq
which is just the standard Newtonian orbit solution
\beq
u_N(\phi)=\frac{1}{\rho}(1+e\cos\phi),~~{\rm where}~~e=u_0\rho
\eeq
where the subscript $N$ refers to the Newtonian solution, $u_0$ is an initial condition constant and $\rho=\ell^2/\alpha m$ is the usual value.

The next step is to now substitute $u\to u_N+\delta u$ into eq.~\ref{eq:u alice} where we only keep one order in perturbation theory. Since $u_N$ part of this expression cancels the usual part of the differential equation from Newton's law, we are left with a differential equation for the perturbation $\delta u$:
\bea
\frac{d^2\delta u}{d\phi^2}+\delta u & = & \frac{3}{\rho}\ralice^2 u_N^2(\phi) \\
& = & \frac{3}{\rho^3}\ralice^2 (1+2e\cos\phi+e^2\cos^2\phi)
\eea
To obtain the complete solution we need to solve for $\delta u$. 
The theory of ordinary differential equations tells us that all we need is {\it any} particular solution, and here is one:
\beq
\delta u = \frac{3}{\rho^3}\ralice^2 \left( 1+e\phi \sin\phi+\frac{e^2}{3}\cos 2\phi +e^2\sin^2\phi\right)
\eeq

The perihelians of the orbit can be obtained by solving for $\phi$ in 
\beq
\rho\frac{du}{d\phi}=-e\sin\phi+3\frac{\ralice^2}{\rho^2}\left(e\sin\phi+e\phi\cos\phi-\frac{2}{3}e^2\sin 2\phi+2e^2\sin\phi\cos\phi\right) =0
\eeq
The existence of the $\phi\cos\phi$ term in this equation, which came from the $\phi\sin\phi$ term in  $\delta u$, is causing the perihelion on the next cycle to shift away from $2\pi$. Defining $\phi=2\pi+\delta$ we can solve for $\delta$ in the perturbative expansion:
\beq
\delta =6\pi\frac{\ralice^2}{\rho^2}.
\eeq

Given our previous computations, we are now able to evaluate the relationship between the extra length scale $\ralice$ and the perihelion advance of an orbit.  In one case, if we have made a measurement of the perihelion advance, we can derive what value $\ralice$ must be to reproduce that value
\bea
\ralice & = & (7.58\times 10^{6}\, {\rm meters})\left( \frac{\rho}{\rm au}\right) 
\left( \frac{T_{\rm orbit}}{\rm years}\right)^{1/2}
\left(\frac{\delta/T_{\rm orbit}}{{\rm arcsec/century}}\right)^{1/2} 
\eea

On the other hand, if we have a theory for what $\ralice$ should be, we can make a prediction for the perihelion advance in units of arc seconds per century:
\beq
\frac{\delta}{T}=\frac{6\pi \ralice^2}{\rho^2 T_{\rm orbit}}
=(1.74 \, {\rm arcsec}\cdot {\rm century}^{-1})\left(\frac{\ralice}{10^7\, {\rm meters}}\right)^2\left( \frac{1\, {\rm au}}{\rho}\right)^2\left( \frac{1\, {\rm yr}}{T_{\rm orbit}}\right).
\eeq

 \section{Philosophical Challenges to Newton's Theory}
 \label{sec:philosophy}

We pause here to describe some foundational questions that Newton's theory faced. There are three main philosophical problems: (1) What is the nature of absolute time and space, and is it necessary to invoke it? (2) Why should the gravitational mass be equal to the inertial mass in the equations of motion? And (3) how does nature enable action at a distance responses? 

Regarding Absolute Space and Time, Newton sets forth his ideas in the first Scholium of {\it Principia}. Almost immediately upon the publication of his book, Newton faced criticism from noted physicists and mathematicians. The most famous adversary regarding this was Leibnitz, who claimed that the only thing that need be talked about, and which ultimately defines space and time, is the relative motions of objects (relativism). Appeals to absolutes make no sense. Newton's friend Samuel Clarke argued vociferously for the absolute viewpoint (substantivalism). Their correspondences are famous, and illuminating in the history of science. Over time these  discussions progressed from what some might think is word quibbling to important physics principles emphasized by Mach and Einstein to name just two.  Pedantic rigor of thinking can lead to the thought processes that generate significantly better theories, and this philosophical problem is arguably an illustration of that.

The second problem, why is the gravitational mass equal to the inertial mass in my mind is the problem that should have kept everyone sleepless for those many centuries when there was not an answer. Newton's theory has nothing to say on the matter, except {\it well, there it is}. These masses are two separate beasts, and why they should be the same? The resolution of this issue is one of the core motivating principles behind General Relativity,  which succeeds in giving a deeper explanation for this curious equality.

The third philosophical problem is sometimes called the problem of action at a distance. There are two aspects of action at a distance. The first is why should two bodies far removed from each other with nothing in between them feel gravitational attraction. Should not there be some ``touching" or medium that carries the gravity force from one body to the other? This action at a distance occurs between particles separated by a large vacuum of nothing. This is hard to take.  Even Newton was disturbed by it, especially the latter aspect. In 1693 he wrote his friend Richard Bentley 
``It is inconceivable that inanimate brute matter should, without the mediation of something else which is not material, operate upon and affect other matter without mutual contact ...."~\cite{NewtonThayer}.

The second aspect of the problem, which is related to the first, is how can two bodies far removed from each other in space instantaneously feel the effect of another's gravitational force. Newton's theory implicitly assumes that all particles feel all other particles' gravitational attraction strength by the exact separations of those particles at each moment of time.  If a particle moves just a little, everybody knows about it instantly and the resolution of forces are adjusted instantly.  To Newton and others, action at a distance was intolerable, but the Newtonian system was the best thing going, and it had tremendous practical value, so it was not to be abandoned despite its flaws.

The issue of instantaneity was noted from the start, and Laplace touched upon it in his highly influential {\it Trait\'e de M\'ecanique C\'eleste}, published from 1797-1825.  He stated that instantaneous propagation did not appear convincing\footnote{``La propagation instantan\'ee qu'ils supposaient \`a cette force me parut peu vraisemblable"\cite{Laplace:1805}.}, and noted that Bernoulli had suspicions as well. But Laplace knew that if the propagation were indeed finite it would have to be extraordinarily fast, and even suggested, incorrectly as it turns out, that some observations imply that it is eight million times that of light.  Laplace briefly brought up the possibility of modifying the inverse square law based on this potential objection but ultimately dismissed it, stating that the simplicity of Newton's theory authorizes us to think of it as a rigorous law of nature\footnote{``En g\'en\'eral, on verra dans le cours de cet ouvrage que la loi de la gravitation r\'eciproque au carr\'e des distances repr\'esente avec une extr\^eme pr\'ecision toutes les in\'egalit\'es observ\'ees des mouvement c\'elestes: cet accord, joint \`a la simplicit\'e de cette loi, nous autorise \`a penser qu'elle est rigoureusement celle de la nature"\cite{Laplace:1805}.}.

Nevertheless, the philosophical challenges to Newton's theory are enough to realize that it was not a complete theory. As we say often in physics today, there must be ``physics beyond the Standard Model". How might signal of ``new physics" show up beyond Newton's theory?  
Let us consider, for example, the disturbing underlying assumption of action at a distance. As we implied above, there are two different issues with action at a distance. There is the aspect of reaching across the mediumless vacuum, and there is the aspect of instantaneous transmission of information to all particles in the universe when one particle moves. 

Transforming our theory from reaching across the vacuum action at a distance to action by local contact is the subject of the theory of fields.  Particles source fields that permeate  spacetime, and other particles experience those fields. Thus, action at a distance is replaced by particle-field interactions in this classical point of view.  The emanating field propagates at finite velocity, which is incorporated self-consistently into modern field theories, retaining causality and introducing the more acceptable action by local contact. 

We do not need to fast forward all the way to the field theories of today to ask how Newton's theory can be pressured experimentally by applying our philosophical worries. The most obvious way one should have thought to do it is by testing the instantaneous aspect of action at a distance.  If one doubts that it is to be rigorously upheld, then we should expect that a quick movement of a body in a mechanical system might yield unexpected results since it might be significantly displaced from its original position by the time the other bodies ``get word" of its flight, and it becomes ambiguous to know what direction and magnitude of force should be applied at all times. Thus, at some sufficiently high speed we might expect to see something unusual -- something unplanned for in the Newtonian world. The trouble is, we do not know a priori what speed this breakdown would occur, and we certainly do not know what new description would be applicable.  

In circumstances like this, it is often best to write down effective theories that satisfy the symmetries of your worldview and do precision measurements to find deviations.  The pattern of deviations or the values of couplings in the effective theory can lead to new insight when explained by a deeper theory.  Bob's $1/r^2$ correction theory and Alice's $1/r^3$ correction theory to the gravity potential in the preceding sections do precisely that. They are Galilean invariant, and satisfy all the symmetries cherished even then: rotational invariance and translation invariance.

We apply this approach of writing down corrections to planetary motion because this is our greatest hope to find cracks in the old classical world view.  Since no cherished symmetries are violated by the additional terms we have found before, we may even expect to find breakdowns of Newton's theory by the orbits of the planets, especially since they are accessible and moving faster with respect to each other and the sun than any laboratory system that could have possible been created on the earth at the time.  Precision measurements of fast planetary motions thus had good reason to be the first place to find breakdown of Newton's theory. No planet moves faster than Mercury.  Indeed, it is Mercury where the first fissures arise,  as we shall describe in the following sections.

\section{Effective Theories}

It is my contention that the concepts of Effective Theories, if understood and held by the early Newtonian scientists, would have led to a prediction that there must {\it necessarily} be an anomalous perihelion precession of Mercury and other planets, and that even the order of magnitude could have been guessed well before Le Verrier's announcement in 1859.  There was no barrier to adopting these ideas in Newton's day, as it requires no new special experimental knowledge, nor knowledge of Einstein's relativity, but rather a more mature approach to how we think about the laws of nature.  In order to present this viewpoint, I shall first give a pr\'ecis of the modern notions of Effective Theories.

At its core, the term Effective Theory is short for a body of evidence that has led us to understand that ``everything depends on everything else" may be true in principle but certainly not true in practice. In a restricted domain, the theory manifests symmetries and properties that provide the ability to calculate observables without the requirement of making reference to features outside the domain.  A simple example of this is that we can compute the trajectory of a football to any practical precision without needing to know the location of Uranus.  The effects of Uranus on the trajectory are suppressed by a relative factor of $\frac{m_e r_e^2}{m_Ud_U^2}\sim 3\times 10^{-14}$, where $r_e$ is the radius of the earth, $d_U$ is the distance from Uranus, and $m_e$ ($m_U$) is the mass of the earth (Uranus). This is much too small to take into account for any practical need.  The diminishing effect of Uranus as $d_U\to \infty$ is the principle of decoupling, which is at the core of Effective Theory utility and is the central reason why science works and we are able to compute and predict observables.

A central concept of Effective Theory is the recognition that a full theory with heavy and light degrees of freedom can be written at low energies in terms of just light degrees of freedom after ``integrating out" the heavy states or ``coarse graining" over the small scales.  We use ``heavy" and ``light" abstractly here, as it could refer to masses, momenta, velocity, etc. The chiral lagrangian of QCD, the Fermi theory of electroweak interactions, the Landau-Ginzburg theory of superconductivity~\cite{Polchinski:1992ed} can all be recognized as an Effective Theory of a more fundamental theory.  

This top-down approach to understanding Effective Theories can give us a multitude of theoretical insights into the nature of simplified low-energy theories.  It is this top-down approach that is traditionally how the power of Effective Theory concepts is demonstrated in particle physics~\cite{Cohen:1993, Rothstein:2003}, fluid mechanics~\cite{Delgado:2005}, material science~\cite{Abrams:2005}, and essentially any other field that has a separation of scales.  However, when considering theories from bottom up,  the concepts we learn from Effective Theories can help us deduce modifications and additions to our present theories that can be tested by experiment. Success then can lead to motivations for inducing a more fundamental theory that reproduces the Effective Theory when restricted to its domain. It is this direction in theory analysis that I emphasize here for our present purposes.

The insight that I would like to focus on, which I believe is the most powerful one when it comes to divining additions and modifications to theories, is the role that symmetries and naturalness play in the construction of the ``complete" Effective Theory.  A symmetry is a recognition that something (a triangle, an equation, etc.) stays the same even if you make a closed set of transformations (i.e., group operations) on that object (rotations by 180 degrees, interchange of $x$ and $y$ variables, etc.). All of our fundamental theories have inherent recognized symmetries in them. We cannot proceed without these recognitions in the Effective Theory, because even the names we give to objects are merely shorthand notation for their symmetry properties (e.g., electrons are spin-1/2 representations of the Lorentz Group with additional gauge symmetry representation labels). 

One of the principle consequences of the Effective Theory approach to establishing natural law is that all possible interactions (or ``terms") consistent with the recognized symmetries of the Effective Theory are generically expected.  There may or may not be additional terms that violate the symmetries, but terms that do not violate the symmetries must be included.  In the realm of Effective Theories within quantum field theory, Weinberg, reflecting on the last three decades of work on the subject, has made the equivalent point that an Effective Theory may be considered self-consistent and not sick ``as long as every term allowed by symmetries is included"\cite{Weinberg:2009bg}.

In short, the precise form of a theory or law is not what is to be taken most seriously -- it is the recognized symmetries.  Upon sorting out the symmetries, the Effective Theory is to be developed with all possible terms consistent with the symmetry, and then qualitative expectations for experiment can be presented.  What remains is measurement and pinning down the actual values of the coefficients to each symmetry preserving interaction term.

\subsection{Application to Newton's Gravitation}

Newton's law of gravitation is that the force between two bodies of masses $m$ and $M$ is inversely proportional to the square of the distance between them, with the proportionality constant being Newton's constant $G$:
\beq
F(r)=\frac{GMm}{r^2},~~{\rm or}~~V(r)=\frac{GMm}{r}
\label{newton law}
\eeq
where $V(r)$ is the potential.
In Book 3 of Principia, Newton states categorically that the inverse square law is ``proved with the greatest exactness from the fact that the aphelia are at rest" and that ``the slightest departure from the ratio of the square would necessarily result in a noticeable motion of the apsides...."~\cite{Principia}.  Thus, the theory was created and solidified as a proposition to the world. 

Newton's inverse-square law was so sacrosanct that few would ever doubt it.  Immanuel Kant in 1747 used the inviability of the inverse-square law to derive that space had three dimensions. This is due to what we would say today is the conservation of gravitational flux lines emanating from a point mass through the surface of a sphere of arbitrary radius.   God could have chosen a different gravity law, Kant says, and the number of spatial dimensions then would have had to be different\footnote{``Zweitens, dass das Ganze, was daher entspringt, verm\"oge dieses Gesetzes [inverse-square law] die Eigenschaft der dreifachen Dimension habe; drittens, dass dieses Gesetz willk\"urlich sei, und da Gott daf\"ur ein anderes, zum Exempel des umgekehrten dreifachen Verh\"altnisses [i.e., inverse-cube law], h\"atte w\"ahlen k\"onnen; dass endlich viertens aus einem andern Gesetze auch eine Ausdehnung von andern Eigenschaften und Absmessungen geflossen w\"are" (section \S 10 in~\cite{Kant1747}).}. This rigid adherence to ``god-given" specific law is ultimately incorrect reasoning, and it is in conflict with modern views of Effective Theories.

The modern sensibility says that we should focus more on the symmetries, and then refashion the complete Effective Theory using them.  What are the symmetries of the Newtonian world? The symmetries are that the laws of physics cannot be affected by one's orientation in space, by one's location in space, nor by one's location in time. The laws must be invariant to any transformation of rotation, spatial translation, or time translation.  These symmetry properties go under the name of Galilean invariance.  As a side comment, the Lorentz invariance of Einstein's special relativity asymptotes to Galilean invariance in the low velocity limit (i.e., when $v\ll c$).

The interaction term of eq.~\ref{newton law} is merely one term in an infinite number of terms that could be written down that are completely consistent with Galilean invariance. An Effective Theory approach would be to introduce them all and investigate the consequences.  There is no meaningful symmetry that demands only the inverse square law interaction. Assured of this, one example would be to embellish Newton's law by
\beq
V_{ET}(r)=\frac{GMm}{r}\left[ 1+\sum_{n=1}^\infty \lambda_n \left(\frac{r_0}{r}\right)^n\right]
+\cdots
\eeq
where $r_0$ is some dimensionful Effective Theory length scale and $\lambda_n$ are dimensionless coefficients, which together with $r_0$ can be found by performing precise experiments. We should note that there are an infinite variety of other terms that could be added, including $r^j$ and $\dot r^k$ interactions, but we streamline the argument by looking only at one class of corrections that decouple as $r\to \infty$. 

\subsection{Inevitable Perihelion Precession}

An extremely important conclusion can already be presented from the rules of Effective Theories. Any deviation from the pure inverse square law will lead to a perihelion precession of the planets, and as the constructed Effective Theory demands additions to the inverse square law there will be an anomalous perihelion precession of the planets.  On the other hand, we know that the inverse square law is approximately correct and thus we have added terms that decouple as $r\gg r_0$.  The perihelion precession of Mercury is very small, and so we expect that $r_0$ should be much less than the orbital radius of Mercury around the sun.  In that case, we are justified in looking at the first-order corrected potential, which we can write as ($\lambda_1r_0\to R$):
\beq
V_{1}(r)=\frac{GMm}{r}\left( 1+\frac{R}{r}\right).
\label{V1 potential}
\eeq
By these arguments of Effective Theory, an anomalous perihelion precession of Mercury is inevitable. It is only a question of what value does $R$ take, which then sets the numerical value of the precession. In the subsequent sections we discuss some arguments for what $R$ might be, from the vantage point of pre-special relativity and pre-general relativity days, and make rough quantitative predictions for the precession rate.

Up to this point we have argued that the focus should have been more on the symmetries of the gravitational theory rather than the concretization of the theory. A more complete Effective Theory for Newtonian gravity would have been accepted and one would have fully expected anomalous perihelion precessions of the planets. A potential similar in form to eq.~\ref{V1 potential} would have been put forward, and the task of theoretically divining or experimentally measuring $R$ would have been the consuming activity.

\section{Mercury's anomalous perihelion precession}

Let us imagine that Bob and Alice are two physicists who are working in the post Le Verrier and pre Einstein era. They are smitten by the Newtonian worldview. They do not wish to do radical things to explain this perihelion precession. They are well-versed in the concepts of Galilean Invariance, Hamilton's Principle, and have an inkling of the ideas of effective theories. Naturally, they  want to describe this precession through a Galilean invariant effective theory of gravity.  Bob announces that he wishes to add a $1/r^2$ correction to the lagrangian. Not wanted to follow in Bob's footsteps, Alice declares that the force law should be even powers of $1/r^{2}$ and so her first correction to the lagrangian is $1/r^3$.  The two lagrangians are
\bea
L_{\rm bob} & = & \frac{1}{2}m\dot\bvec r^2+\frac{\alpha}{r}\left( 1+\frac{\rbob}{r}\right) \\
L_{\rm alice} & = &  \frac{1}{2}m\dot\bvec r^2+\frac{\alpha}{r}\left( 1+\frac{\ralice^2}{r^2}\right) 
\eea
where $\alpha=GMm$, with $G$ being Newton's constant, $M$ is the mass of the sun, and $m$ is the mass of the planet under consideration. These are the two lagrangians of Bob and Alice that we studied in a previous lecture.  These new laws of Bob and Alice require the introduction of a new fundamental length scale $R_i$. They do not know what that length scale is, but they have hopes that the new data will pin it down for them.

Before we look more closely at Bob and Alice's theories, we should remark again that in the classical history of gravity, there were early attempts to explain anomalies by changing Newton's laws, even in the manner of Alice and Bob. Such theories go under the name of ``Clairaut laws". Clairaut proposed in 1745 that Newton's law should be corrected by a $1/r^4$ force term in order to explain some thought-to-be anomalies in the movement of the lunar perigee. However, he found in the end there was not a discrepancy, which buried such laws deeper into the dustbin of history. Newcomb commented in 1882 that such laws were ``out of the question" because they disrupted the gravitational strength so wildly at very close distances where the correction term would come to dominate~\cite{Newcomb1882}.  As late as 1910 Newcomb, the world's leader on this issue, was stating that all the data up to that point ``... seems to preclude the possibility of any deviation from that law [Newton's inverse-square law]" and that Mercury's perihelion advance is best explained by ``the hypothesis of Seeliger"~\cite{Newcomb1910}, which was a zodiacal light theory that contained intra-Mercurial distributions of orbital matter minimally disruptive to all other astronomical observations except Mercury's perihelion advance (see, e.g., chapter 4 of ~\cite{Roseveare:1982}).  

Bob and Alice's theory are a return to the Clairaut law in some ways. In the next few subsections we merely state the effect they would have on planetary orbits.  After a discussion of Effective Theories and how they apply to this problem, we shall proceed with a somewhat fanciful alternative history of how deviations from Newton's laws could have been explained and interpreted from the point of view of Effective Theories after the anomaly was announced by Le Verrier. But it should be kept in mind, and will be emphasized again in the concluding section, that these theories could have been anticipated, and perhaps even should have been anticipated, before Le Verrier's announcement.  

\subsection{Analyzing Bob's $1/r^2$ Correction Theory}

From eq.~\ref{deltatorbitbob} we can compute in Bob's theory that it is necessary that $\rbob=4.4\, {\rm km}$ if Mercury is to have the measured 43 seconds of arc per century in its perihelion precession. Given this value of $\rbob$, Bob can make predictions for the perihelion advance of other planets. Using eq.~\ref{eq:deltarbob} he finds $\delta/T_{\rm orbit}=8.6''$ of arc per century for Venus's perihelion precession and $3.8''$ for the earth. These predicted values compare favorably to the measurements for Venus and Earth presented in Table~\ref{Planet Precessions}. The predictions are well within the errors, and Bob is pleased because he has found a way to explain the anomaly while yet retaining Galilean invariance as a fundamental symmetry of spacetime. He has done this through the means of a simple expansion correction to Newton's law of gravity. Nothing radical was done.

\begin{table}[t] 
\centering 
\begin{tabular}{cc}
\hline\hline
Planet & $\delta/T$ (arcsec/century) \\
\hline
Mercury & $43.11 \pm 0.45$ \\
Venus & $8.4\pm 4.8$ \\
Earth & $5.0\pm 1.2$ \\
\hline\hline
\end{tabular} \caption{Anomalous perihelion precession rates of the planets compared to expectations from Newton's law of gravity and taking into account all other sources of precession (effects of other planets orbits, etc.)~\cite{PerihelionData}. More modern references test gravity (including precession rates) through the parameters of the so-called parametrized post-Newtonian (PPN) approach~\cite{Will:2005va}.}
\label{Planet Precessions}
\end{table}

Despite the successes, Bob is not totally satisfied. He wants to know if he can argue for this new length constant in nature $\rbob$. It's a very strange distance $4.4\, {\rm km}$.  He wonders how he can formulate this distance from all the invariants swirling around him. It should not depend on the mass of each planet, he reasons, because we have just shown that one value of $\rbob$ appears to work universally well for all planets.  The other options we have to build a length scale are from Newton's constant $G$, the mass of the Sun $M$ and angular momentum. Bob fails to find any natural combination that will give $4.4$ km. 

Before giving up he recalls that his intuition has told him that there is some characteristic high speed such that Newton's simple laws become strained (see sec.~\ref{sec:philosophy}). He does not know what that speed value is, and his new law is just as much action at a distance as the old one, but he carries on by giving this new speed a name, $\vbob$. With this new undetermined speed in hand he realizes immediately that he can form a new length scale $GM/\vbob^2$. Can this be the origin of $\rbob$?  What value must $\vbob$ be to recover $\rbob=4.4$ km? A simple calculation yields
\beq
\vbob=\sqrt{\frac{GM}{\rbob}}=1.7\times 10^8\, {\rm m/s}.
\eeq

This quantity $\vbob$ that Bob has derived is a very curious number! His colleagues down the hall have been working on the theory of electromagnetic phenomenon and a speed very close to that keeps showing up in their equations, $c=3.0\times 10^8\, {\rm m/s}$.  This is the propagation speed of light. He decides this cannot be a coincidence, but he is not sure what to make of it. He decides to define a new scale based on these thoughts, the ``sun's electro-gravity scale"  $R_{EG}\equiv GM/c^2$. $\rbob$ can now be written in terms of this definite scale $\rbob=\lambda_{\rm bob}R_{EG}$. It is very curious that the data fits very well if $\lambda_{\rm bob}=3$ is an integer. He writes on a piece of paper his new theory of gravity
\beq
L_{bob}  =  \frac{1}{2}m\dot\bvec r^2+\frac{GM m}{r}\left( 1+3\frac{GM/c^2}{r}\right),
\label{eq:bob theory}
\eeq
and he is pleased with its simplicity, elegance and symmetry. He does not know how the speed of light $c$ crept in, but he is satisfied since his lagrangian looks ``natural" given that there are no really big or really small numbers populating it. Furthermore, he knows that if he must construct a new length scale with a speed, the ``natural" next known threshold of speed is the speed of light, and so this correction is ``natural" to explore. He feels he is on to something big.

Bob finds another interesting connection with this scale $GM/c^2$.  He recognizes that there is a small radius $R_E$ of a infinitesimal (i.e., radius less than $R_E$) spherical body of mass $M$ for which an object going the speed of light would not be able to escape. This light-speed trapping radius is a curiosity: if light were corpuscular in any sense, as Newton and others thought it might be, then we could see no light emanating from within the radius $R_E$ of the massive body. This sets a mystery scale to gravity that requires further scrutiny and may be a length scale associated with changes in gravity.  The computation of this scale is simple in the Newtonian world, and is
\beq
R_E=\frac{2GM}{c^2}~~{\rm (light~non-escape~radius)}
\label{first R}
\eeq
This is only a factor of two different than the value of $R$ he has derived from the perihelion precession rate. It should be noted that $R_E$ is the precisely the Schwarzschild radius derived in General Relativity, which is a well-known special scale for spherically symmetric objects for more reasons than just what was stated above~\cite{Schwarzschild:1916,Wald:1984}.
Furthermore, it should be recalled that the speed of light was being quantitatively estimated~\cite{Romer:1676} even before Newton's {\it Principia}, and by 1729 it was known to within a few percent~\cite{Bradley:1729}, and so this scale had precise meaning from the very beginning days of Newtonian gravity.

Despite these interesting connections, Bob gets nervous looking over his equations.  Eq.~\ref{eq:u bob} seems to indicate that if $\rho < 2\rbob=6GM/c^2$, the orbits do not make sense anymore, as the equations formally say $r<0$ which is nonsensical.   He relaxes briefly when he realizes that $2\rbob$ is only 9 km, which is well below the orbital radius of any planet, and furthermore it is even below the radius of the sun, which is $7\times 10^5\, {\rm km}$. Thus, there is no danger that some small object rotating around the sun would have no chance to be described by Bob's theory, since it would be inside the sun.

Nevertheless, he is still a bit uncomfortable. Nowhere in his derivation was the radius of the sun ever required. In principle, all that mass of the sun could have been at one infinitesimal point for all the equations knew. Nevermind how to pack all that mass in with a radius less than 9 km, it is a possibility in principle that such a tightly packed object exists, and if it did, there is no way his theory could describe close-by orbits with characteristic orbital latus rectum size $\rho<9\, {\rm km}$.  He knows his theory cannot be the end all of all the theories anyway due to not knowing why $c$ crept into his equations, despite that being the natural next ``speed scale" to consider, but now he is even more discomfited because he can imagine configurations where his theory just cannot even give an answer.  But that is for another day. He has succeeding in explaining the precessions of Mercury, Venus and Earth and that is enough for a day's work.  And that is what Effective Theories do. They explain the day's work -- Bob clearly has made progress -- but there is more to be learned and understood. Effective Theory practitioners understand that all possible questions cannot be resolved instantly, and that there are necessarily deeper effective theories to come.

\subsection{Analyzing Alice's $1/r^3$ Correction Theory}

Alice now wishes to make definite her lagrangian with $1/r^3$ potential corrections  by specifying the value of $\ralice$ from Mercury's anomalous perihelion precession and then predicting what the  other precession rates are.   Upon fitting Mercury data she finds $\ralice=9.04\times 10^5\, {\rm meters}$. Using that fixed value for all planets she then predicts $\delta /T_{\rm orbit}=4.43''$ of arc per century for Venus and $1.4''$ of arc per century for the Earth. The Venus result is nearly $2\sigma$ off compared to the measurement, and the Earth result is about $3\sigma$ off of the measurement (see Table~\ref{Planet Precessions}). Alice has a choice now. She can say her theory predicts that further refined measurements of the precession rates will yield smaller central values of the precession rates for Venus and Earth in concert with her theory. Or, she can take the $3\sigma$ discrepancy seriously and attempt to modify her theory.

Alice makes the right choice and seeks to modify the theory.  She computes what $\ralice$ needs to be for each planetary case to precisely hit the measured values. She finds 
\beq
\ralice^{\rm mercury}=90\times 10^7\, {\rm m},~~\ralice^{\rm venus}=1.3\times 10^7\, {\rm m},
~~\ralice^{\rm earth}=1.5\times 10^7\, {\rm m}.
\label{eq:alice lengths}
\eeq
 Similar to Bob, she begins to think about how these length scales can be identified with all the quantities that she has available to her in the problem: $M$, $m_{\rm planet}$, and $\ell$.
She cannot come to a satisfactory answer. These constants alone are not enough to form the length scales of eq.~\ref{eq:alice lengths}. 

However, in Alice's trials she notices something interesting. The $\ralice$ lengths are proportional to angular momentum divided by mass of the planet, $\ralice^i\propto \ell_i/m_i$, with the same proportionality constant. This constant has the dimensions of an inverse velocity. She decides to call it $v_{\rm alice}$ and solves for its value:
\beq
\ralice^i=v^{-1}_{\rm alice}\frac{\ell_i}{m_i}~~\Longrightarrow~~
v_{\rm alice}=\frac{\ell_i}{m_i\ralice^i}=3.0\times 10^8\, {\rm m/s}
\eeq
 Alice also has colleagues that work on electromagnetism and she recognizes this value as exactly the speed of light, $v_{\rm alice}=c$. How did that happen? She does not know, but she is surely excited about the result, as she too recognises that $c$ is the next fundamental ``speed threshold" and so is a ``natural" value in the Effective Theory correction. She has explained all the planetary precession data. She writes down on a piece of paper her new theory of gravity,
 \beq
 L_{\rm alice} =  \frac{1}{2}m\dot\bvec r^2+\frac{GMm}{r}\left( 1+\frac{1}{c^2}\frac{\ell^2/m^2}{r^2}\right).
 \label{eq:alice theory}
 \eeq
which like Bob's theory possesses symmetry and has a measure of elegance and simplicity.
  
 As she reflects on her theory she realizes that since angular momentum is $\ell \sim mrv$, where $v$ is the velocity of the planet orbiting the sun, the second term inside the parenthesis can be thought of as an $m$-independent $v^2/c^2$ correction to the Newtonian gravitational potential. Thus, she believes that she will be the first to show that the simple inverse-square law of Newton is corrected by factors of $v^2/c^2$. As the speed of the planet gets closer to the speed of light, Newton's theory begins to crack. So far the basic assumptions of spacetime symmetries -- Galilean Invariance -- are not breaking down, just the simple form of Newton's theory of gravity. Despite these successes of her theory, she remains slightly dissatisfied with one aspect. How can she convince herself, much less others, that her theory is better than Bob's? Surely one or the other or some combination of these corrections are required by nature, she reasons, but can they be determined from deeper theory principles?  The answer is yes, and Einstein's General Relativity is that theory.
 
\subsection{Gerber's ``Utterly Worthless" Theory}

Before going to Einstein's General Relativity, let us comment briefly on velocity dependent approaches to augmenting Newton's law. Manipulations of the Newtonian potential were initiated in earnest well after Laplace's work with the goal of rigorously incorporating finite speed effects of gravity. The most straightforward approaches failed. However, Paul Gerber proposed in 1898~\cite{Gerber:1898} a velocity dependent potential correction that correctly accounted for Mercury's perihelion precession:
 \beq
 V(r,v)=-\frac{M}{r}\left( 1-\frac{v}{c}\right)^{-2}
 \eeq
 where $c=3\times 10^8\, {\rm m/s}$ is the speed of light, and $v$ is the velocity of Mercury in the Sun-Mercury center of mass system.  
 
 Gerber's theory captured the attention of many due to its combined simplicity and effectiveness in accommodating Mercury's anomalous perihelion precession rate. For example, Mach wrote, ``Only Paul Gerber [reference to 1898 paper] studying the motion of Mercury's perihelion ... did find that the speed of propagation of gravitation is the same as the speed of light"\cite{Mach:1901}.  He was attacked for not giving good reasons for his theory -- a topic we shall take up below -- but he did provide a simple theory that worked. It was also a ``natural" theory due to its utilization of $c$ as the next fundamental speed scale of the theory.
 
 Seventeen years after Gerber's potential, the question of Mercury's perihelion precession was resolved powerfully by Einstein's GR~\cite{Wald:1984}. At low velocities the first-order correction to gravitational attraction of Gerber's theory matches the first-order correction of Einstein's theory. However, Einstein's approach had coherent principles and unassailable logic, and thoughts about Gerber's theory quickly faded away.

Despite the success in accommodating Mercury's perihelion precession, Gerber was roundly criticized for his theory.  The strength of the reaction that Gerber faced seems harsh for somebody who actually did write down a simple theory of no free parameters with the speed of light in it that worked. It is as though the deep thinkers at the time knew there was something appealing about Gerber's work, but could not quite put their finger on it, and so harshly criticized it as a community building exercise to dismiss that kind of apparently principle-less approach to physics.
 
 Einstein, commenting on Gerber's theory well after he had developed his own theory of General Relativity summarized the attitudes well: ``But specialists in the field agree not only that Gerber's derivation is thoroughly incorrect, but that the formula cannot even be obtained as a consequence of Gerber's leading assumptions. Mr. Gerber's paper is therefore {\it utterly worthless}"\cite{Capria:1999} (italics are mine). This appears to be an overly strong dismissal of Gerber's simple theory that gained so much attention. 
 
 Pauli, in his famous Encyclopedia article on Relativity said,
 \begin{quote}
 Recently, an earlier attempt by P. Gerber has been discussed which tries to explain the perihelion advance of Mercury with the help of the finite velocity of propagation of gravitation, but which must be considered {\it completely unsuccessfully} from a theoretical point of view. For while it leads admittedly to the correct formula -- though on the basis of false deductions -- it must be stressed that, even so, only the numerical factor was new.~\cite{Pauli81} (italics mine)
 \end{quote}
 
 Whatever can be said of Gerber and his theory and the faulty logic behind his theory, it was not ``utterly worthless" or ``completely unsuccessful".  I believe it was a crude attempt at effective theory analysis. It was something he may have intuited but was unsuccessful in articulating well due to the mindset and style of physics of the day.  Back then, no term was allowed to augment a theory without it being derived first from a deeper principle. The standard rigor of the day was that laws were exact by argument and deduction, and any deviations or changes must be accounted for by a replacing new principle.
 
 An excellent example of this prevailing attitude is provided by Max Born in his book on Einstein's theory of relativity~\cite{Born24}. He describes briefly the case of Mercury's anomalous perihelion precession and then goes on to harangue all those people before Einstein who generated ad hoc solutions to the problem:
\begin{quote}
Changes in the laws [Newton's laws] have been proposed, but they have been invented quite arbitrarily and can be tested by no other facts, and their correctness is not proved by accounting for the motion of Mercury's perihelion. If Newton's theory really requires a refinement we must demand that it emanate, without the introduction of arbitrary constants, from a principle that is superior to the existing doctrine in generality and intrinsic probability. Einstein was the first to succeed in doing this.
\end{quote}
This attitude is partially in conflict with our understanding of Effective Theories today.  The introduction of arbitrary constants is a key step in the construction of Effective Theories, and the role of experiment is to pin those down.  If anything, the ad hoc inventors of changes in Newton's law were too sheepish about introducing arbitrary parameters, and instead got tangled up with incoherent ``deep reasons" for their particular laws. 

Effective Theory is an intermediate step between an old regime (e.g., Newton's laws) and a new regime (e.g., Einstein's General Relativity), and this intermediate step necessarily has ``arbitrary couplings" and does not ``emanate from a principle that is superior to the existing doctrine". Instead, it says that the existing doctrine should be taken to its utmost seriousness (e.g., Galilean invariance) and data should fit the parameters of all allowed interactions, and perhaps a deeper new theory can come along later to explain the relations among those parameters.
 
Although Gerber's theory was not worthless, it is not as valuable as Einstein's General Relativity.  Alice and Bob's effective theories would not have been worthless either had they written it down much earlier. They would have been an intermediate stepping stone from one principled theory to the next that would have predicted the existence of Mercury's perihelion precession and motivated earlier discovery of the phenomena.

 \section{Perturbation from General Relativity}
 
We have talked about Einstein's General Relativity being the deeper theory that explains Mercury's perihelion precession. It is worthwhile in these lectures to go through that computation to see how it comes about.

We wish to compute the trajectory of a particle subject to a central, radially symmetric gravitating source in the general approach followed, for example, by~\cite{Hartl2003}. The metric applicable for this computation is the Schwarzschild metric:
\beq
ds^2=-\eta(r)c^2dt^2+\frac{dr^2}{\eta(r)}+r^2d\theta^2+r^2\sin^2\theta d\phi^2
\eeq
where 
\beq
\eta(r)=1-\frac{2GM}{c^2r}=1-\frac{r_s}{r},~~{\rm where}~~r_s\equiv 2GM/c^2
\eeq
The quantity $r_s$ is the Schwarzschild radius. This defines the metric tensor to be
\beq
g_{\alpha\beta}={\rm diag}(-\eta(r),\eta(r)^{-1},r^2,r^2\sin^2\theta)
\eeq
in the $(t,r,\theta,\phi)$ basis. 
Note that the signature of the metric (asymptotically weak field far away) in normal rectilinear coordinates is $g^{\alpha\beta}={\rm diag}(-1,1,1,1)$.

The Schwarzschild metric is unperturbed by making shifts in the time direction and by making shifts in the angular direction $\phi$. These define Killing vectors $\xitime^\lambda=(1,0,0,0)$ and $\xirot^\lambda=(0,0,0,1)$. The nice property of a Killing vector is that when dotted into the four-velocity vector $dx^\alpha/d\tau$ the result must be constant along the geodesic motion: 
\beq
\xi^\lambda \frac{dx_\lambda}{d\tau}=g_{\alpha\beta}\xi^\alpha \frac{dx^\beta}{d\tau}={\rm const}.
\eeq
Applying this theorem to the Schwarzschild metric gives
\bea
g_{\alpha\beta}\xitime^\alpha \frac{dx^\beta}{d\tau} & = & \eta(r)\frac{dt}{d\tau}=c_1 \\
g_{\alpha\beta}\xirot^\alpha \frac{dx^\beta}{d\tau} & = & r^2\sin^2\theta \frac{d\phi}{d\tau}=c_2 
\label{eq:x1}
\eea
where $c_1$ and $c_2$ are mere constants. We know that independence of time implies conservation of energy, and we also know that independence of rotation implies conservation of angular momentum. Thus, we know that $c_1$ is some function of energy, and we know that $c_2$ is some function of angular momentum as we usually define the quantities. However, at this stage we do not know the precise correspondence, so it is prudent to just carry the constants $c_1$ and $c_2$ with us until the precise relations become obvious.

From eq.~\ref{eq:x1} we solve for $dt/d\tau=c_1/\eta(r)$ and $d\phi/d\tau=c_2/r^2\sin^2\theta$. Now, we should simplify this all by taking the orbit in the $\theta=\pi/2$ plane and so $d\phi/d\tau=c_2/r^2$. Please note, conservation laws have given us this, and this is where deep physics lies. Now, let's expand out  the defining equation of the four-velocity 
\beq
g_{\alpha\beta}\frac{dx^\alpha}{d\tau} \frac{dx^\beta}{d\tau}=-1,~~{\rm which~gives}
\eeq
\beq
-\eta(r)\left(\dtdtau\right)^2+\frac{1}{\eta(r)}\left(\drdtau\right)^2+r^2\left(\dphidtau\right)^2=-1
\eeq
for the Schwarzschild metric.
Substituting the values of $d\phi/d\tau$ and $dt/d\tau$ that we obtained above from the Killing equations, we find
\beq
-\frac{c_1^2}{\eta(r)}+\frac{1}{\eta(r)}\left(\drdtau\right)^2+\frac{c_2^2}{r^2}=-1
\eeq
After carrying out some algebra one finds
\beq
\frac{m c^2}{2}(c_1^2-1)=\frac{1}{2}mc^2\left( \frac{dr}{d\tau}\right)^2-\frac{GMm}{r}
+\frac{mc^2 c_2^2}{2r^2}-\frac{GMmc^2_2}{r^3}
\label{eq:x2}
\eeq

The form of eq.~\ref{eq:x2} is very suggestive of our equation for energy of a particle in an orbit, and the correspondence becomes precise if we make the identifications
\beq
\frac{mc^2}{2}(c^2_1-1)\equiv E ~~~{\rm and}~~~c_2^2\equiv \frac{\ell^2}{m^2c^2}.
\eeq
We also can identify $\tau=ct$ in the non-relativistic limit.  It turns out that this substitution is acceptable for the problem at hand as long as $\dot r\ll r\dot\phi$, which is generally the situation for low eccentricity orbits, and certainly the case for the planetary orbits of our solar system.  Making these identifications the energy equation becomes
\beq
E=\frac{1}{2}m\left(\frac{dr}{dt}\right)^2+\frac{\ell^2}{2mr^2}-\frac{GMm}{r}\left(1 + \frac{\ell^2/m^2c^2}{r^2}\right).
\eeq
This is the energy equation for a particle in Newtonian gravity except for the small shift in the effective potential
\beq
\Delta V_{eff}(r)=-\frac{GMm}{r}\left( \frac{\ell^2/m^2c^2}{r^2}\right)
\eeq
which is precisely the same correction to Newton's  theory we derived earlier from Alice's effective theory approach to explain Mercury's precesion in eq.~\ref{eq:alice theory}. 

There are multiple ways to derive the correction to Newton's gravity law for the particular problem of perihelion precessions. In our derivation, we found Alice's theory correction.  This is also the result derived in General Relativity by many other authors (see e.g., \cite{Schutz90,Goldstein02,Hartl2003}).  However, another approach to the General Relativity derivation gives Bob's theory, and that has been demonstrated by a set of different authors (see e.g., \cite{Pauli81,LandauB,Iwasaki71,Donoghue:2009mn}).  These two theories, if treated as god-given complete theories, are not equivalent. However, they are equivalent results for {this problem} as all approximations and culling of the General Relativity  terms have been carried out with  the sole purpose of finding the perihelion precession.  In the end, the precession rate angle per orbit period from either correction is the same:
\beq
\delta = \frac{6\pi GM/c^2}{a(1-e^2)}
\eeq
Algebraically, the orbital identity 
\beq
\ell^2=GMm^2a(1-e^2).
\eeq
is what guarantees that the two solutions predict the same anomalous perihelion precession rate.
So, we see that Albert explains both Alice's theory and Bob's theory, and puts them on firmer footing.

\section{Conclusions}

At the beginning of these lectures we decided that Newton's law of Gravitation was very successful in describing the orbits, but that it is not the precise law that captures our most profound admiration. Rather, it is the symmetries that the theory possesses. We elevated those to the highest principles and constructed reasonable effective theories that could be expected by the data. We illustrated the results with two theories: Bob's $1/r^2$ and Alice's $1/r^3$ potential correction theories. Both theories were able to account for the perihelion rate naturally. We even made the case that philosophical challenges to Newton's world view, if taken seriously, could presage the size of Mercury's correction that was actually measured by Le Verrier. This is done with the aid of ``naturalness" arguments about the speed of light being the next speed scale of nature by which to construct corrections to Newton's potential. In this way the concepts of natural effective theory have some predictive power. That power is certainly qualitative, but also to some degree quantitative. 

Einstein had keen insights into the nature of space and time and developed the theory of General Relativity based on them. It describes gravity at a deeper level, and one of its first orders of business was to compute the anomalous precession rate of Mercury to see if it could account for the discrepancy between Newton's theory and measurement.  The answer is yes, and we have shown that this correction matches nicely the effective theory results of Alice and Bob.

Einstein's General Relativity theory is  ``better" than Alice's theory or Bob's theory for two reasons. First, it gives a deeper principles understanding of the correction with no additional free parameters. This deeper understanding is nothing other than further assumptions on spacetime symmetries that panned out.  Second, Einstein's theory is a more complete theory of gravity that makes additional predictions (such as bending of light, and binary pulsar spin-down) that are confirmed by data. Alice or Bob's theory clearly cannot match the riches of General Relativity and so cannot be considered as fundamental as Einstein's.  

Despite Bob and Alice's theory coming up short, the general lesson remains. Newton's theory was an effective theory, which is in some aspects superceded in success by Bob and Alice's effective theory, and Bob and Alice's effective theories are superceded in success by Einstein's General Relativity. The obvious next question is whether Einstein's General Relativity theory can be succeeded in success by another theory. A deeper theory that perhaps could be explained as effective theory expansion of Einstein's theory for the purposes of solving some lower energy precision measurement problem. There is little doubt that is the case~\cite{Donoghue:1994dn}.  

Finally, one of the most profound shifts in our thinking over the decades, illustrated well by the Perihelion precession example, is that it is really no longer appropriate to speak of ``the correct theory." There is no correct theory. Our tasks are to improve theories via the effective theory approach, to seek deeper and simplifying assumptions that account for it, solidify those into a new theory, and then treat that new theory as an effective theory, and repeat. These steps are accomplished by continually improving and refining observations and theory computations that enable us to choose between effective theories, followed by deducing deeper new symmetries that force its inevitability.  Theories are never to be trusted -- they are always ``wrong" in the end -- and with concerted effort we can even anticipate when and how they will break down. 

The concepts of Effective Theory lead one to predict qualitatively that a perihelion precession of Mercury was {\it a priori} guaranteed even knowing only the experimental facts of the Newtonian era. In particular, elevating symmetries above the concretization of hypothesized law, in this case the rigid devotion to the inverse square law, is the basic ingredient that would have led unambiguously to this conclusion. The general approach to science during the Newtonian era required almost complete devotion to concrete laws and their propositional justifications, which impeded its progress toward developing theory enhancements guided by symmetries and naturalness. Gerber, a school teacher who was perhaps not as indoctrinated in this more rigid fashion, found a potential that worked yet then made unjustified arguments for why it should be true. Effective Theories give the best of both words: deep but modest justifications for theories that can anticipate data and fit the data.

We have also shown that even during the time of Newton a reasonably well supported hypothesis for the perihelion precession of Mercury could have been put forth that is close to the actual experimental result of $43''$ of arc per century.  This is a clear illustration of how the ideas of Effective Theory can be utilized to extrapolate modestly beyond the rigidly set forth laws of fundamental physics.

\noindent
{\it Acknowledgments:} I wish to thank D. Baker, J. Kumar, S. Martin and B. Skow for helpful discussions.


\end{document}